\documentclass[sn-mathphys]{sn-jnl}

\usepackage{graphicx}        
\usepackage{multirow}        
\usepackage{subfigure}       
\usepackage{url}             
\usepackage[numbers,sort&compress]{natbib}  
\usepackage{amsmath,amssymb,amsfonts}      
\usepackage{algorithmic}     
\usepackage{textcomp}        
\usepackage[x11names,dvipsnames]{xcolor}   
\usepackage{hyperref}        
\usepackage{array}           
\usepackage{tabularx}        
\usepackage{listings}        
\usepackage{tcolorbox}       
\usepackage[numbers,sort&compress]{natbib}
\usepackage{xcolor}
\usepackage{amsthm}
\usepackage{booktabs}
\newtheorem{definition}{Definition}

\lstdefinelanguage{yaml}{
  keywords={true,false,null,y,n},
  sensitive=false,
  comment=[l]{\#},
  morecomment=[l]{\#},
  morestring=[b]',
  morestring=[b]",
}

\lstdefinestyle{yamlstyle}{
  language=yaml,
  basicstyle=\scriptsize\ttfamily,
  keywordstyle=\color{violet}\bfseries,
  commentstyle=\color{Gray}\itshape,
  stringstyle=\color{ForestGreen},
  showstringspaces=false,
  breaklines=true,
  tabsize=2,
  morekeywords={
  version,description,
  inputs,outputs,steps,
  id,name,image,command,
  type,message,action,
  onError,
  schema_file,data_file,
  raw_data,matched_data,filtered_data,test_result
 },
  moredelim=**[il][\color{Gray}]{\#}
}

\lstdefinestyle{arazzostyle}{
  language=yaml,
  showstringspaces=false,
  basicstyle=\ttfamily\scriptsize,
  keywordstyle=\color{violet}\bfseries,
  commentstyle=\color{gray},
  stringstyle=\color{teal},
  tabsize=2,
  breaklines=true,
  morekeywords={
    title, version, arazzo,info,workflows,steps,inputs,outputs,x-policy, x-compliance,file,
    operationId,parameters,onError,description,
    policyRefs,verifiedBy,enforcement,
    rules,type,article,requirement,evidence,
    id,name,engine,source
  },
  morecomment=[l]{\#}
}

\begin{document}

\title{Compliance Management for Federated Data Processing}

\author[1]{\fnm{Natallia} \sur{Kokash}}\email{nkokash@uva.nl}

\author[1]{\fnm{Adam} \sur{Belloum} }

\author[1]{\fnm{Paola} \sur{Grosso} }

\affil[1]{\orgdiv{Institute of Informatics}, \orgname{University of Amsterdam}, \country{The Netherlands}}

\abstract{
Federated data processing (FDP) offers a promising approach for enabling collaborative analysis of sensitive data without centralizing raw datasets. However, real-world adoption remains limited due to the complexity of managing heterogeneous access policies, regulatory requirements, and long-running workflows across organizational boundaries. In this paper, we present a framework for compliance-aware FDP that integrates policy-as-code, workflow orchestration, and large language model (LLM)–assisted compliance management. 
Through the implemented prototype, we show how legal and organizational requirements can be collected and translated into machine-actionable policies in FDP networks. 
}


\maketitle

\section{Introduction}

In the era of ML and AI, few organizations have sufficient data to conduct cutting-edge research alone, yet sharing data across institutions raises significant privacy and security concerns. Federated data processing (FDP) is a data management paradigm in which data are accessed and processed across multiple independent sources without physically consolidating them~\cite{Calvanese2018,Lenzerini2002}.

Despite growing adoption, no unified guidelines exist for deploying and managing federated workflows~\cite{Naeem2020}. Practical implementation faces two compounding challenges. First, the heterogeneity of available tools~\cite{liu2024survey} forces collaborating consortia to negotiate not only technical choices but also governance rules and processes. Second, current legal frameworks are built around the assumption of data sharing rather than data-local processing~\cite{dataguidance2025,dlapiper2025dataprotection}, leaving no standardized, programmable governance systems that new collaborations can adopt off-the-shelf.

To address this gap, we propose an architecture for setting up FDP workflows that manages sensitive data in accordance with relevant data protection regulations. Its core execution layer is Brane~\cite{valkering2021}, a secure, container-based environment for privacy-aware data access. Brane alone, however, is insufficient: compliant FDP also requires tooling to elicit regulatory constraints, translate them into software policies and monitoring objectives, implement and test workflows, align ML algorithms with site-specific data representations, and manage project assets after execution.

We therefore introduce \emph{BraneHub}, a coordination hub for collaboration setup, policy specification, dynamic workflow adaptation, deployment, and monitoring. Compliance responsibility is divided between BraneHub — which describes FDP projects and manages onboarding of data providers and consumers — and the Brane environment, which enforces programmable access and purpose control over containerized data processing functions. Although BraneHub is designed to configure Brane networks, its template-based approach and service composition stack are applicable to other federated virtual network types.

To reduce the burden of compliance engineering, BraneHub offers a low-code FL deployment pipeline~\cite{zhuang2022easyfl} integrating large language models (LLMs) and retrieval-augmented generation (RAG). Guided by structured questionnaires capturing contextual factors — data access roles, access duration, jurisdiction, and purpose — the system assists researchers with limited coding experience in drafting project-specific terms, policy specifications, and monitoring rules, and in authoring machine-level access control policies in OPA/Rego~\cite{opa_playground} or eFLINT~\cite{VANBINSBERGEN2022140}. Participants can join projects via self-service registration, submit constraints, and receive updates through BraneHub. 

The remainder of this paper is organized as follows. Section~\ref{sect:brane} introduces Brane and its privacy-aware processing paradigm. Section~\ref{sect:scenario} clarifies FDP network requirements through a healthcare scenario. Section~\ref{sect:approach} presents our approach to FDP project setup with attention to compliance management. Section~\ref{sect:discussion} reflects on solution readiness and evaluation. Section~\ref{sect:related-work} discusses related work, and Section~\ref{sect:conclusions} concludes with future directions.

\section{Brane FDP Networks}
\label{sect:brane}

Brane~\cite{valkering2021} is a lightweight yet expressive framework for multi-site scientific applications via programmable orchestration, built around three interlocking capabilities: (i) user-friendly workflow representation, (ii) encapsulation via container packaging, and (iii) automated cross-site execution — together ensuring flexibility, portability, and usability across distributed, heterogeneous environments.

Orchestration logic is expressed through a domain-specific language (DSL) that abstracts infrastructure details (sites, container runtimes, scheduling), allowing domain scientists to define execution pipelines without managing underlying systems. Functionality is encapsulated into self-contained container images via one of Brane's four package builders, providing portability, reproducibility, and dependency management across sites.

Most existing workflow management systems were designed for a single administrative domain. Cluster-centric schedulers such as Slurm~\cite{Yoo2003Slurm} and PBS/Torque~\cite{PBSPro} assume shared storage within a single cluster. Cloud-native engines such as Google Cloud Composer~\cite{CloudComposer} or Azure Logic Apps~\cite{AzureLogicApps} are bound to a single vendor and require custom integration for hybrid deployments. General-purpose orchestrators such as Luigi~\cite{Luigi} and Airflow~\cite{Airflow} support heterogeneous backends but through explicit task definitions rather than transparent site switching. Workflow engines with multi-backend support such as Nextflow~\cite{DiTommaso2017Nextflow} and Snakemake~\cite{Koster2012Snakemake}  allow backend selection via configuration, but each execution still targets a single environment; native cross-site execution requires additional federation layers. Crucially, none of these systems provide built-in policy enforcement for federated data governance or support data-local execution across independent administrative domains.

Brane addresses these limitations by routing tasks, distributing data, and coordinating execution across heterogeneous multi-site platforms. Workflows are represented as graphs of tasks and data dependencies; the orchestrator communicates with domains via metadata-only control messages. Sensitive data remain under local domain control and are subject to site-specific access policies. Domains are autonomous and may refuse or constrain any orchestrator-requested data transfer.

A key feature introduced in the EPI project~\cite{alsayedkassem} is Brane's policy-driven approach to trustworthy data processing. Data owners retain control through operational policies enforced via eFLINT~\cite{10.1145/3425898.3426958} — a formal policy specification language — and through dynamic deployment of virtual network functions (VNFs) for low-level network security. OPA/Rego support as a runtime policy checker is under active development~\cite{policyreasoner+opa}.

\begin{figure}
    \centering
    \includegraphics[width=0.9\textwidth]{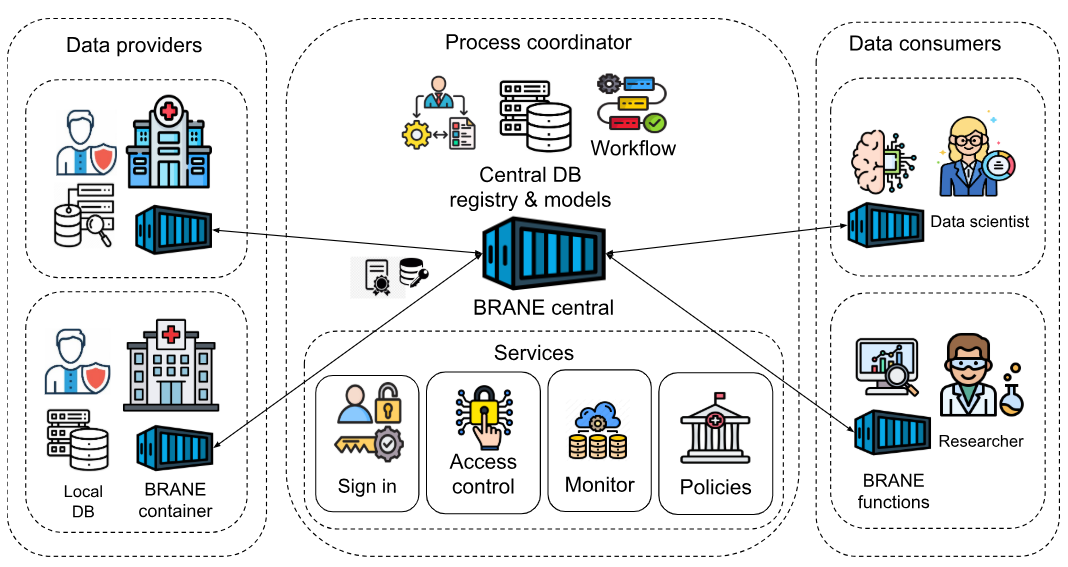}
    \caption{Brane FDP network}
    \label{fig:brane-diagram}
\end{figure}

Figure~\ref{fig:brane-diagram} illustrates the high-level architecture of the Brane-centered FDP network setup. In this setup, data providers maintain local databases and deploy Brane containers that execute approved workflows within institutional boundaries, ensuring that raw data remains local. Brane central maintains a registry of models and metadata, orchestrates distributed workflows, and enforces governance through authentication, access control, monitoring, and policy services. Data consumers, such as data scientists and researchers, interact exclusively with Brane central to submit analytical workflows and retrieve authorized results. The workflow engine coordinates secure task execution across participating providers, aggregates permitted outputs, and collects audit trails.

\section{Running Scenario}
\label{sect:scenario}

We consider a motivating scenario to elicit requirements for cross-organizational collaboration involving sensitive data. 

A research and development (R\&D) facility within a pharmaceutical company or research institute conducts early-stage studies on a target condition as part of a long-term drug development pipeline. Investment at this stage is substantial, with end-to-end development costs ranging from several hundred million to multiple billions of euros depending on trial scale, disease rarity, and regulatory burden \cite{wouters2020estimated}. To assess prospective global demand before committing to downstream clinical trials, such facilities seek information on relevant patient cohorts, including prevalence, geographic distribution, and contextual factors associated with onset. Where legally permitted, this can include selected patient medical history, laboratory records, imaging data, or, in later phases, identifiable personal data to recruit eligible participants.

\begin{tcolorbox}[colback=gray!5!white, colframe=gray!75!black, title=Scenario Summary, sharp corners, boxrule=0.8pt]
\scriptsize
The facility gathers information about patient cohorts, such as condition frequency and circumstances of onset. Where permitted, this includes access to medical history, laboratory tests, and images for evaluation of PK/PD models, AI methods, or digital twins. In later research stages, personal data may be collected for patient recruitment in clinical trials.
\end{tcolorbox}

For example, developing a drug targeting pediatric hydrocephalus requires epidemiological data on incidence in children aged 2 to 8 months, including maternal medical history, underlying health conditions, medication or substance use, and pregnancy-related complications. Hydrocephalus affects approximately 1–2 per 1,000 live births globally~\cite{Dewan2018HydrocephalusEpidemiology}, yet publicly available pediatric hydrocephalus imaging datasets remain very limited~\cite{Xu2025HyKid}.

Initial feasibility studies typically last 6–12 months and involve a small multidisciplinary team (clinical, R\&D, epidemiology, health economics). Full development programs involve multiple functional teams:
\begin{itemize}
    \item \textit{Discovery \& preclinical}: medicinal and computational chemists, pharmacologists, toxicologists.
    \item \textit{Clinical development}: research scientists, biostatisticians, trial managers, data managers, pharmacovigilance officers, and regulatory specialists.
    \item \textit{External partners}: contract research organizations (CROs) handling trial monitoring, patient recruitment, and data collection.
    \item \textit{Cross-functional governance}: legal/compliance, manufacturing, and commercial planning teams.
\end{itemize}

The governance of pediatric and maternal health data, relevant for our scenario, operates within a multilayered regulatory framework that combines national and regional data protection laws, pediatric-specific safeguards such as age-dependent consent requirements and parental authorization, institutional policies, and ethics/IRB approvals. Particularly sensitive categories—including genetic, maternal, and reproductive data—are subject to heightened protections, compliance with Good Clinical Practice guidelines~\cite{ICH2025GCP}, and restrictions on secondary use.

Across jurisdictions, data protection regimes consistently establish core obligations governing the collection, processing, and sharing of personal data (see Table~\ref{tab:regs}). These include requirements for lawful consent, purpose limitation, data minimization, security safeguards, transparency, and enforceable data subject rights, as summarized in Table~\ref{tab:regs-categories}.

\begin{table*}[ht]
\caption{Regulatory frameworks for data privacy}
\label{tab:regs}
\scriptsize
\centering
\begin{tabular}{|p{1.6cm}|p{2.2cm}|p{3.8cm}|p{3.8cm}|}
\hline
\textbf{Region} & \textbf{Regulatory Framework} & \textbf{Scope} & \textbf{Patient Data Rights} \\
\hline
USA & \textbf{NIST (e.g., SP 800-162)} & Cybersecurity guidance; ABAC access control policies & Not regulated directly; focuses on access control \\
\hline
USA & \textbf{HIPAA} & Health data privacy and security & Access and consent rights \\
\hline
EU & \textbf{GDPR} & Personal data protection & Full data subject rights \\
\hline
EU & \textbf{ePrivacy Directive (2002/58/EC)} & Privacy in electronic communications (metadata, cookies) & Informed consent; confidentiality and protection of communications \\
\hline
Netherlands & \textbf{NEN 7510} & Healthcare information security & Aligned with GDPR \\
\hline
Germany & \textbf{BSI IT-Grundschutz} & IT security for critical infrastructures & Enforced via GDPR and national privacy laws \\
\hline
UK & \textbf{DSP Toolkit} & NHS data protection & Based on GDPR \\
\hline
Australia & \textbf{Privacy Act 1988}, \textbf{My Health Records Act 2012} & Personal and health data; sector-specific rules for health records & Rights to access, correct, and control health data \\
\hline
Canada & \textbf{PIPEDA}, \textbf{Provincial Health Laws (e.g., PHIPA)} & Personal info in commercial activities; health-specific provincial laws & Access, correction, and consent for health data use \\
\hline
Japan & \textbf{APPI} & Personal info, with special rules for sensitive/medical data & Rights to be informed, access, correct, or delete data \\
\hline
China & \textbf{PIPL} & Personal information processing & Rights: access, correction, deletion, portability, consent \\
\hline
South Korea & \textbf{PIPA} & Collection and processing of personal data & Rights to access, correct, delete, or suspend processing \\
\hline
\end{tabular}
\end{table*}

\begin{table*}
\caption{Categories of compliance constraints regulating data privacy}
\label{tab:regs-categories}
\scriptsize
\centering
    \begin{tabular}{|p{2.6cm}|p{6cm}|p{3.8cm}|}
    \hline
    \textbf{Category} & \textbf{Description} & \textbf{Examples / References} \\
    \hline
    Access Control Policies & Limits access based on roles, attributes, or purpose. & {HIPAA Access Control}, {ABAC} \\
    \hline
    User Consent & Requires informed, explicit, and revocable consent from the data subject. & {GDPR Article 7}, {HIPAA Authorizations} \\
    \hline
    Data Minimization & Collect/share only necessary data. & {GDPR Article 5(1)(c)} \\
    \hline
    Data Localization & Requires data to be stored or processed within specific geographic boundaries. & {EU Data Governance Act}, {China PIPL} \\
    \hline
    Anonymization \& Pseudonymization & Requires removal or obfuscation of identifiers before data sharing. & {GDPR Recital 26}, {HIPAA De-identification} \\
    \hline
    Purpose Limitation & Data may only be used for explicitly stated purposes. & {GDPR Article 5(1)(b)} \\
    \hline
    Data Integrity \& Accuracy & Requires data to be accurate and kept up-to-date. & {GDPR Article 5(1)(d)} \\
    \hline
    Auditability \& Transparency & Requires systems to log access and processing. & {GDPR Article 30}, {HIPAA Audit Controls} \\
    \hline
    Security Safeguards & Technical/organizational measures like encryption, firewalls, access control, etc. & {NIST SP 800-53}, {HIPAA Security Rule} \\
    \hline
    Data Subject Rights & Includes rights to access, correct, port, or erase data. & {GDPR Articles 15--22}, {HIPAA Right to Access} \\
    \hline
    Third-party Data Sharing & Limits sharing with external entities without agreements or risk assessments. &
    {GDPR Article 28}, Data Use Agreements \\
    \hline
    \end{tabular}
\end{table*}

\subsection{Implications for Software Infrastructure Requirements}

In such complex FDP scenarios, multiple regulatory instruments and contractual agreements define which data may be accessed, by whom, and under which conditions. Consequently, FDP deployment requires integrated software subsystems to operationalize these requirements. Core components include:

\begin{enumerate}
    \item \textit{Compliance management}: trace consent conditions, purpose limitation, data access retention, and contractual constraints across federated participants.
    \item \textit{Secure data access}: encrypted communication, role- and attribute-based access control.
    \item \textit{De-identification and pseudonymization}: privacy-preserving transformations with controlled re-linkability.
    \item \textit{Data harmonization and interoperability}: common data models, metadata standards, and semantic mappings for cross-site analyses without centralizing raw data.
    \item \textit{Access logging, monitoring, and audit}: recording and evaluating data access and computation for accountability and compliance verification.
\end{enumerate}

\begin{figure}
    \centering
    \includegraphics[width=.9\linewidth]{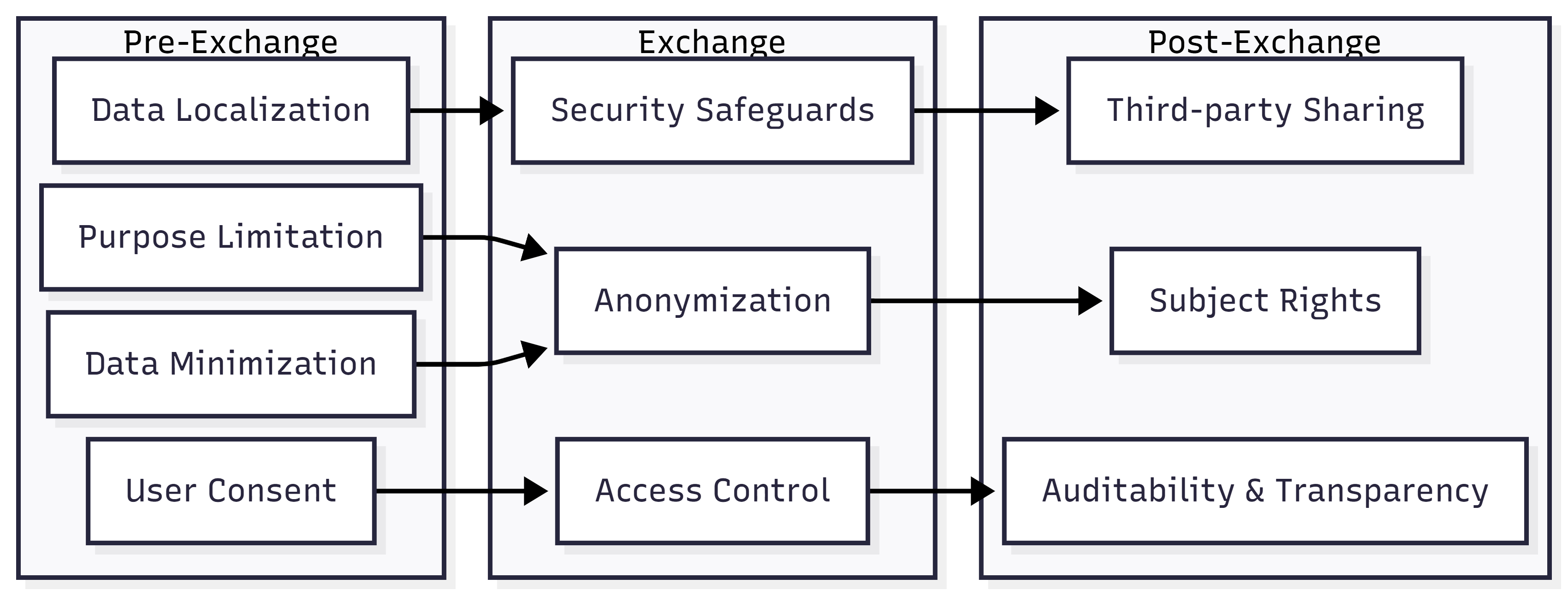}
    \caption{Preserving data privacy through the processing cycle}
    \label{fig:data-stages}
\end{figure}

A framework providing access to sensitive data must ensure compliance throughout the project lifecycle. Figure~\ref{fig:data-stages} shows the stages of privacy-aware data exchange—pre-exchange, exchange, and post-exchange—with key protective tasks at each stage. 

While Brane addresses some requirements directly (II and V) or via policies and VNFs (III and IV~\cite{kokash2025ontology}), it lacks full governance and traceability from legal requirements. Its academic-style policy-based access control requires formal logic expertise, creating a barrier for domain experts, data stewards, or compliance officers. Manual translation of legal/organizational requirements into formal rules quickly becomes impractical.

We therefore propose an infrastructure integrating a design-time collaborative setup and project management platform with Brane-centered workflow execution. This setup supports longevity, transparency, testability, maintainability, and scalability, leveraging LLMs to translate natural-language policies into machine-readable specifications while tracing links between legal requirements and derived policies.

\begin{figure*}
    \centering
    \includegraphics[width=.99\linewidth]{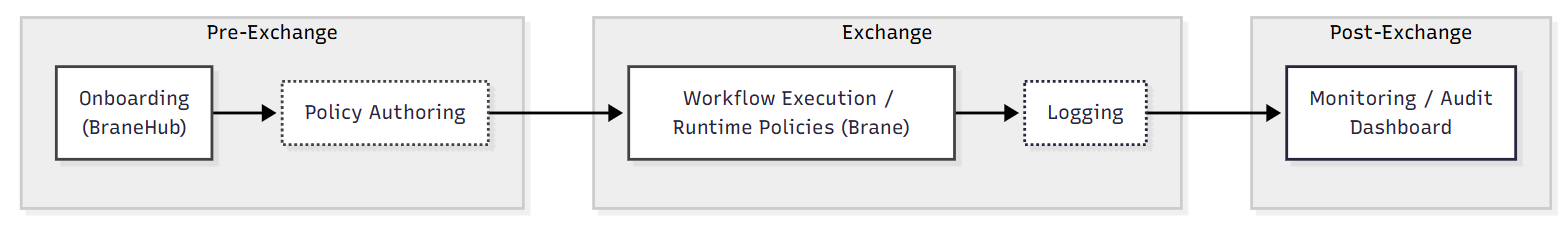}
    \caption{Data-privacy protection via policy management in Brane-centered FDP}
    \label{fig:stages-brane}
\end{figure*}

\section{Overview of the FDP framework}
\label{sect:approach}

\subsection{Management Portal and Project Setup}

\begin{figure*}
    \centering
    \includegraphics[width=0.9\textwidth]{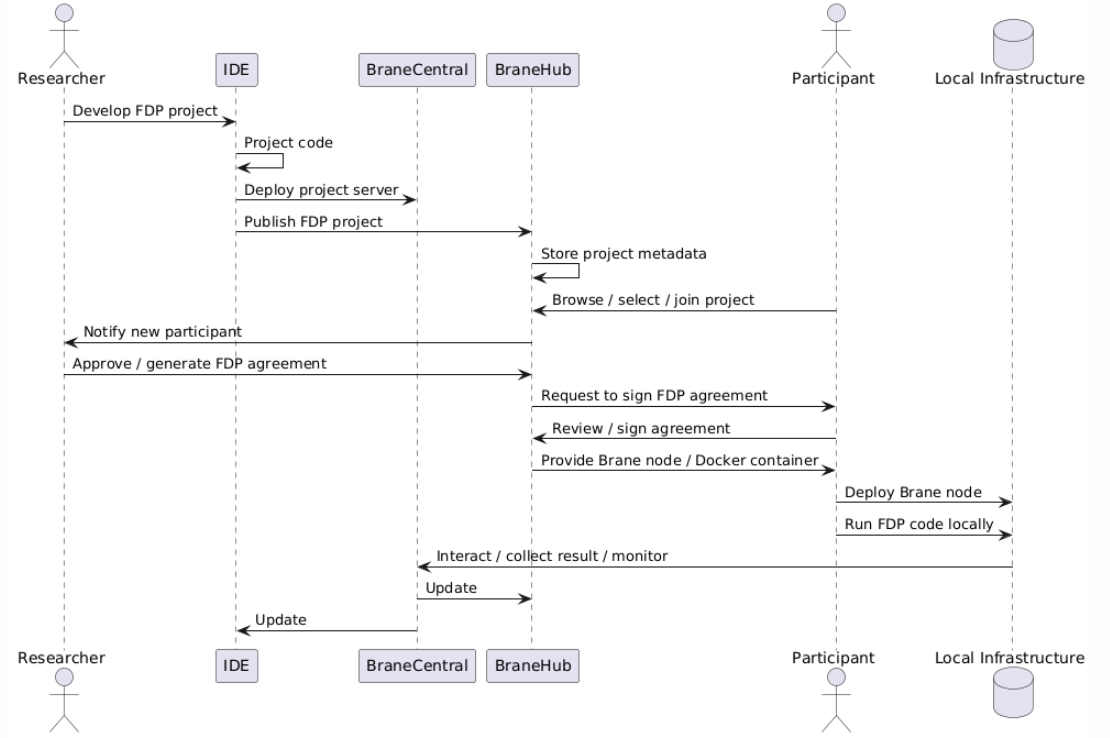}
    \caption{Dynamic FDP setup}
    \label{fig:collaboration-setup}
\end{figure*}

The BraneHub portal serves as the central collaboration and onboarding interface for participants in the FDP project, facilitating structured interaction between data consumers and data providers. Through the portal, data consumers specify their expected datasets and analytical requirements, while simultaneously gaining insight into the data assets and constraints available at participating institutions. Figure~\ref{fig:collaboration-setup} presents a sequence diagram of this setup process, highlighting how the portal supports recruitment, requirement elicitation, and semantic harmonization to bridge the gap between consumer needs and provider capabilities, thereby enabling compliant joint processing.
\begin{enumerate}
    \item Researcher starts from developing a project locally using mock-up data and deploys it on the Brane infrastructure in a test mode.
    \item The testing setup will allow the developer to supply mock-data via the nodes simulating participants to make sure the data processing is robust and is working as expected.
    \item Once the data processing pipeline is ready, the researcher deploys it in the production mode and publishes the project (study goals, call for required data, incentive for participants, data access restrictions to activate the recruitment process).
    \item BraneHub deploys a Central Brane node that will coordinate tasks in the project workflow.
    \item Participants browse/select projects and submit requests to join the project using a questionnaire that inquires about the data they possess and are willing to grant access to. BraneHub notifies the researcher about a participant's request.
    \item The researcher analyses the request, if necessary, contacts the participant to clarify the information, customizes the FDP workflow configuration, and approves or rejects the participant. Integrated AI assistant and/or policy evaluation agent can be engaged to automate this process. 
    \item BraneHub notifies the participant and provides a link to install a preconfigured container (Brane node).
    \item BraneHub data use agreement is generated and offered (in global study it can depend on regional location of the participant, to guarantee region-relevant rights).
    \item The access and harmonization of technical data has to be agreed upon. Compliance enforcement depends on the data the participant grants access to.
\end{enumerate}

Data collection pipeline for our running scenario looks as follows: 

\begin{tcolorbox}[colback=gray!5!white, colframe=gray!75!black, title=Deployment prompt, sharp corners, boxrule=0.8pt]
\scriptsize
I am a data analyst designing an application to learn from federated data. The application needs two inputs:
\begin{itemize}
    \item an expected data schema supplied by me (data consumer)
    \item a data file in CSV format supplied by a data provider
\end{itemize}
I am not allowed to see the data but can perform a statistical test and read back the result. My data processing pipeline includes the following functions:  \emph{readData}, \emph{matchToSchema}, \emph{filterData}, \emph{performTest}. The \emph{matchToSchema} operation may fail if the data format is incompatible. Write a YAML code to package this pipeline to a container that the data provider will execute on their machine behind the firewall and send me back the test result.
\end{tcolorbox}

Given a configuration file, the workflow designer uses Brane to compile a Brane package and BraneCli tool to deploy packages in containers. The new package can be tested locally or using mock clients' test nodes (Brane nodes deployed on various machines to ensure cross-platform interoperability) before it is pushed to the remote organization's node to process actual data.   

\subsection{LLM-assisted Policy Authoring}

The LLM-assistant within the BraneHub platform can suggest the list of frameworks based on the study objectives. For example, Table~\ref{tab:regs} lists prominent legal frameworks that regulate and protect healthcare data worldwide. 
As Table~\ref{tab:regs-categories} shows, a number of data protection measures are prescribed for any system dealing with patient data. Such repeated compliance obligations must be supported by a FDP framework systematically and used by default to govern workflows orchestrated by the framework. However, the combination of relevant regulatory requirements is specific to the application and depends on the context defined by all stakeholders. Once the documents relevant for the context are established, they are integrated into the vector store for conversion to system requirements.

The \emph{Researcher Interface} (Table~\ref{tab:researcher-interface}) and \emph{Participant Interface} (Table~\ref{tab:participant-interface}) are implemented as configurable web templates generated from structured questionnaires governed by JSON schemas. These schemas formally define required fields, validation rules, and domain-specific constraints, allowing the portal to adapt to diverse project types (e.g., clinical trials, observational studies, federated AI training) and application domains beyond healthcare. Through the Researcher Interface, investigators register study purpose, expected datasets, model classes, processing logic, and ethical approvals, thereby operationalizing purpose limitation, data minimization, and legal basis documentation. The Participant Interface enables healthcare institutions and other data providers to describe available datasets, technical capabilities, internal governance policies, and contractual requirements, supporting organizational due diligence on access control, localization, and security safeguards. Together, these schema-driven interfaces systematically capture the information required to reconcile semantic differences and regulatory expectations across stakeholders. With the support of LLM-based agents, even informally expressed collaboration terms articulated in natural language through the BraneHub can be structured, formalized, and translated into machine-interpretable, workflow-ready specifications.

\begin{table*}
\caption{Researcher Interface (Project Registration Form)}
\label{tab:researcher-interface}
\centering
\scriptsize
\begin{tabular}{l l p{7cm}}
\textbf{Field Label} & \textbf{Input Type} & \textbf{Description} \\
Project title & Text input & Title of the FDP project. \\
Research institution & Text input & Name of the institution conducting the study. \\
Contact email & Email input & Primary contact email for correspondence. \\
Study objective & Text area & Purpose of the data processing. \\
Data required & Multi-select & Examples include imaging, EHR, genomics (can be used to enforce Data Minimization principles). \\
Data sensitivity level & Radio button & Low / Medium / High. Determines consent requirements or additional safeguards. \\
Security measures & Checklist & Encryption, secure containers, access logs, pseudonymization. \\
Result sharing policy & Text area & Specifies whether models or aggregated statistics will be shared externally. \\
Participant responsibilities & Text area & Includes adherence to access policies, logging, and other operational obligations. \\
Legal basis & Dropdown & GDPR, HIPAA, Public Interest, Research with Waiver, etc. \\
Third-party collaboration & Yes/No & If yes, requires appropriate agreements and notifications. \\
\end{tabular}
\end{table*}

\begin{table*}
\caption{Participant Interface (Project Join Form)}
\label{tab:participant-interface}
\centering
\scriptsize
\begin{tabular}{l l p{7cm}}
\textbf{Field Label} & \textbf{Input Type} & \textbf{Description} \\
Organization & Text input & Legal name of the healthcare institution. \\
Contact person & Text input & Designated representative for the FDP project. \\
Contact email & Email input & Official contact email for project communications. \\
Location & Dropdown & Country/region of data storage to assess localization constraints. \\
Data available & Multi-select & Types of data available (e.g., imaging, EHR, lab results). \\
Data sharing & Text area & Institution-specific legal or policy-based constraints. \\
Consent mechanism & Yes/No & Indicates if patient consent has been obtained for this data use. \\
Internal review & Yes/No & Whether an internal ethics or compliance review is mandatory. \\
Other clauses & Text area & Clauses required by the institution for participation. \\
\end{tabular}
\end{table*}

\subsection{Implementation}
\begin{figure*}
    \centering
    \includegraphics[width=1.0\textwidth]{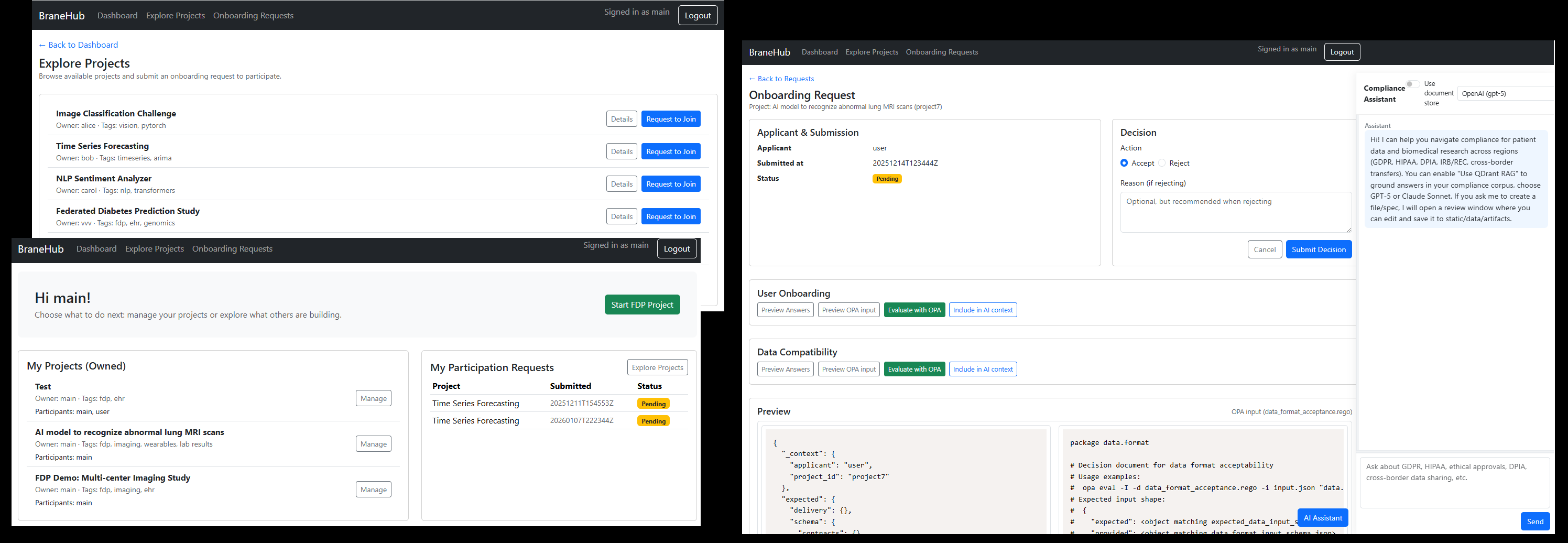}
    \caption{BraneHub, a Web application to manage federated projects}
    \label{fig:branehub-screenshot}
\end{figure*}

Our proof-of-concept implementation of BraneHub~\cite{braneHubGitHub} covers project registration, participant onboarding, and compliance-aware decision support, with a focus on pre-integration phases where technical, legal, and organizational compatibility must be established prior to any data processing. The application comprises the following components:
\begin{itemize}
\item \textbf{Core Application}: A Flask entry point responsible for user authentication, dashboard management, project creation, and onboarding workflow orchestration.
\item \textbf{Domain Services}:
\begin{itemize}
\item \texttt{onboarding.py}: Implements onboarding logic and state management.
\item \texttt{data\_format.py}: Handles validation and formatting of project-specific data requirements.
\end{itemize}
\item \textbf{OPA Integration}: Manages communication with the OPA server and evaluates onboarding requests against Rego policies.
\item \textbf{AI and Vector Search}: Client implementations for Anthropic, OpenAI, and Qdrant, exposing an API for document vectorization and semantic search.
\item \textbf{Policies and Data}: Stores policy files and associated metadata for onboarding and data format validation.
\item \textbf{Frontend}: Jinja2 templates and static assets supporting the web-based user interface.
\end{itemize}
Project owners define FDP projects by specifying objectives, expected data types and formats, sensitivity levels, and governance parameters. Prospective participants submit onboarding requests comprising structured questionnaire responses and machine-readable descriptions of their locally available datasets and interfaces. BraneHub presents these artifacts to the project owner for review, as shown in Figure~\ref{fig:branehub-screenshot}, which illustrates three views: (i) active projects available for the authenticated user to join, (ii) projects owned by that user, and (iii) the details of a submitted onboarding request, including the data specification and the OPA/Rego policy evaluating its compatibility with the expected format.
To formalize compatibility and compliance assessment, BraneHub evaluates owner-defined expectations against participant-provided declarations using OPA and, optionally, LLMs. Structured inputs are assembled automatically and evaluated against purpose-built Rego policies, yielding allow/deny decisions accompanied by explanatory output — unmet requirements, conditions, or advisory notes — without requiring access to the underlying data. Where policy evaluation benefits from contextual reasoning beyond formal rule matching, LLM-based evaluation serves as a complementary mechanism, as illustrated by the region-based acceptance criterion in Listing~\ref{lst:onboarding}.

\lstdefinelanguage{Rego}{
  morekeywords={package,import,default,true,false,if},
  sensitive=true,
  morecomment=[l]{\#},
}

\lstset{
  basicstyle=\scriptsize\ttfamily,
  keywordstyle=\color{blue}\bfseries,
  commentstyle=\color{gray}\itshape,
  stringstyle=\color{teal},
  breaklines=true,
  showstringspaces=false,
  columns=fullflexible,
}

\begin{lstlisting}[language=Rego, caption={Simple onboarding policy}, label={lst:onboarding}]
package region.onboarding

# Accepts a participant if their location is within the given locations list.
# Expected input shape: {
#   "location": "EU", "acceptable_locations": ["EU", "NL", "BE", "DE"...] }

default allow = false

allow if {
  # Ensure the location is present in the acceptable_locations collection
  input.acceptable_locations[_] == input.location
}

# Returns: {"allow": true|false}
decision := {"allow": allow}
\end{lstlisting}
If the web interface permits users to specify their location via a free-text field, legitimate participants may be rejected by the OPA policy when they express their location in syntactically different but semantically equivalent forms (e.g., “Europe,” “European Union,” or the name of an EU member state). To make the decision process more inclusive, the workflow designer would need to enumerate all acceptable variants within the policy, thereby either constructing an exhaustive list of regions and jurisdictions or risking incomplete coverage.

An LLM-based policy checker can mitigate this limitation through the prompt shown in Listing~\ref{lst:prompt}. By leveraging natural language reasoning, the LLM-based onboarding assistant can interpret semantically equivalent inputs without requiring explicit enumeration in the policy. This reduces the burden on data consumers to anticipate all possible formulations during questionnaire design, while enabling data providers to communicate more nuanced information about data characteristics and access constraints. 

\lstdefinelanguage{MyPython}{
  morekeywords={def,return,if,else,for,in,import,from,as,with,True,False,None},
  sensitive=true,
  morecomment=[l]{\#},
  morestring=[b]",
}

\lstset{
  basicstyle=\scriptsize\ttfamily,
  keywordstyle=\color{blue}\bfseries,
  commentstyle=\color{gray}\itshape,
  stringstyle=\color{teal},
  breaklines=true,
  showstringspaces=false,
  columns=fullflexible
}

\begin{lstlisting}[language=MyPython, caption={Policy prompt for region-based onboarding}, label={lst:prompt}]
    input_data = {
        "location": location,
        "acceptable_locations": acceptable_locations
    }
    
    prompt = f"""
    You are a policy evaluation engine. Decide whether to accept a participant based on the following input data: {json.dumps(input_data, indent=2)}

    Rule: Accept a participant (allow: true) if their location is a valid representation or sub-region of one of the acceptable_locations. Otherwise, deny (allow: false).

    Return ONLY a JSON object in the following format:
    {{"allow": true}} or {{"allow": false}}
    """
\end{lstlisting}

BraneHub provides AI-assisted support for interpreting onboarding information and policy outcomes. Users may selectively inject questionnaire fragments, data-format specifications, and policy results into the integrated AI assistant to inquire about implications for the workflow. AI-generated outputs are advisory and can be stored alongside formal policy artifacts to support documentation and traceability. 

\subsection{Workflow specification and policy checkers}

Currently, Brane accepts tasks with a simple ad-hoc workflow specification format that defines inputs, functions to run (steps), and outputs (see an example in Listing~\ref{lst:yaml}). Functions that do not have data dependency are executed concurrently.  
 
\begin{lstlisting}[style=yamlstyle, caption=Federated Analysis YAML Job Specification, label={lst:yaml}]
name: FederatedWorkflow
version: "1.0"
description: Secure workflow to process client data
inputs:
  schema_file:
    type: File
    description: "Expected data schema file provided by the analyst"
  data_file:
    type: File
    description: "Client data file (e.g., CSV format) to be processed securely"
steps:
  - id: readData
    name: Read Data
    image: analyst/federated-stat-test:latest
    command: ["python", "pipeline.py", "readData"]
    inputs: ["data_file"]
    outputs: ["raw_data"]
  - id: matchToSchema
    name: Match to Schema
    command: ["python", "pipeline.py", "matchToSchema"]
    inputs: ["raw_data", "schema_file"]
    outputs: ["matched_data"]
    onError: 
      action: fail
      message: "Schema matching failed."
  - id: filterData
    name: Filter Data
    command: ["python", "pipeline.py", "filterData"]
    inputs: ["matched_data"]
    outputs: ["filtered_data"]
  - id: performTest
    name: Perform Statistical Test
    command: ["python", "pipeline.py", "performTest"]
    inputs: ["filtered_data"]
    outputs: 
      - name: test_result
        type: File
        description: "Result of the statistical test"
outputs:
  test_result:
    type: File
    description: "The result to be sent to the analyst"
\end{lstlisting}
    
This format does not include policy requirements bound to distributed processes~\cite{ramezani2014supporting,lopez2020business}. Policies are pre-loaded or provided in separate files per each node. The following runtime policy reasoners are supported by Brane:

\begin{itemize}
    \item \emph{POSIX} (Portable Operating System Interface) is a IEEE standard that specifies the expected behavior of Unix-like operating systems and their core utilities, including pathname formats and  permission models. The POSIX reasoner within Brane is a tool that does not require any input and simply checks read/write file permissions during workflow execution on data access attempts. 
    \item \emph{eFLINT} is the main reasoner that expects  
    formal rule-based policy definitions grounded in Hohfeld's framework for legal reasoning~\cite{dAlmeida2016}. Its purpose is to express obligations, permissions, prohibitions, and institutional facts to enable reasoning over legal and organizational norms (which is not necessarily useful for the domain scientist interested in deploying a functional FDP workflow for a specific purpose).
    \item \emph{OPA/Rego} is an industry-level reasoner for practical management of access policies, recently integrated with Brane to define policies for data-centered FDP workflows.
\end{itemize}

The \emph{LLM-based reasoner} presented in this work is currently applied only for onboarding (pre-exchange) process, but theoretically can be employed for runtime policy reasoning as well (provided that the security and privacy concerns related to the AI agent use~\cite{lim2022privacy_ai_agents, al-kharusi2024open_source_ai_privacy_security} are properly addressed). 

In future work, we aim to enable the integration of policies with workflow specifications. For example, Arazzo specification~\cite{openapi_arazzo_specification} is a viable candidate for designing a compliance policy definition extension.
The OpenAPI and the Arazzo specifications are complementary standards from the OpenAPI Initiative: 
\begin{itemize}
    \item OpenAPI defines the static surface of an individual API — its endpoints, request and response schemas, authentication mechanisms, and data types — serving as the contract between provider and consumer;
    \item Arazzo extends this foundation by enabling ordered, dependent sequences of calls across one or more OpenAPI-described services, capturing conditional logic, data passing, and failure handling required to express complete business workflows.
\end{itemize}
Where OpenAPI answers what an API can do, Arazzo answers how a set of APIs must be orchestrated to deliver a specific outcome.
In both in OpenAPI and Arazzo, any field prefixed with x- is a specification extension, meaning it is vendor- or implementation-defined and carries no standardized semantics. Our running scenario could look as shown in Listing~\ref{lst:arazzo} (input and output parameters are as in Listing~1 and are skipped for brevity).

\begin{lstlisting}[style=arazzostyle, caption=Federated Workflow with GDPR Consent, label={lst:arazzo}]
arazzo: 1.0.0
info:
  title: FederatedWorkflow
  version: "1.0.0"
  description: Secure workflow to process client data in a federated environment.
x-policy:
  engine: opa
  policyRefs:
    - id: gdpr_rules
      source: https://gdpr-info.eu/
workflows:
  FederatedWorkflow:
    description: Secure federated statistical test workflow with patient consent
    steps:
      - id: readData
        name: Read Data 
        x-compliance:
          rules:
            - id: GDPR-Article5
              description: Raw data not visible to analyst.
              requirement: confidentiality
              verifiedBy: system-policy
              enforcement: true
      - id: matchToSchema
        name: Match to Schema 
        x-compliance:
          rules:
            - id: GDPR-Article25
              description: Keep only schema-matched fields.
              requirement: data_minimization
              verifiedBy: static-analysis
      - id: filterData
        name: Filter Data 
        x-compliance:
          rules:
            - id: GDPR-Article6
              description: Process only consented records.
              requirement: consent
              verifiedBy: opa:gdpr_rules/explicit_consent
              evidence: consent_registry_ref
              enforcement: true
      - id: performTest
        name: Perform Statistical Test
        x-compliance:
          rules:
            - id: GDPR-Article32
              description: Ensure anonymized output.
              requirement: anonymization
              verifiedBy: opa:gdpr_rules/anonymization_check
      - id: failIncompatibleSchema
        name: Handle Schema Error
        description: Stop on schema mismatch.
\end{lstlisting}

\subsection{From Regulatory to System Requirements}

In addition to the compliance requirements addressed via data access policies, the FDP framework must provide system-level functions to fulfill legal obligations. For example, GDPR Article~16 --- \emph{``The data subject shall have the right to obtain from the controller without undue delay the rectification of inaccurate personal data concerning him or her''} --- is operationalized by nine coordination and system requirements summarized in Table~\ref{tab:RQ}.

\begin{table*}[ht]
\centering
\scriptsize
\caption{System Requirements for Rectification of Personal Data}
\label{tab:RQ}
\renewcommand{\arraystretch}{1.4}
\begin{tabularx}{\textwidth}{|p{0.6cm}|>{\raggedright\arraybackslash}p{3.8cm}|X|}
\hline
\textbf{RQ} & \textbf{Functional Area} & \textbf{Requirement Description} \\
\hline
RQ1 & User Request Handling & UI for data subjects to submit rectification requests; each request logged with a unique reference number. \\
\hline
RQ2 & Identity Verification & Use multi-factor authorization, email confirmation, or ID validation for processing rectification requests. \\
\hline
RQ3 & Data Rectification Processing & Authorized personnel or automated systems must correct data without delay; changes must be tracked. \\
\hline
RQ4 & Automated Accuracy Checks & Validate inputs via format checks, consistency, or external data to reduce errors. \\
\hline
RQ5 & Data Synchronization and Propagation & Ensure rectified data is updated across all relevant systems and services. \\
\hline
RQ6 & Response Time and Compliance Tracking & Enforce rectification deadlines and log processing time for compliance. \\
\hline
RQ7 & Audit Logging and Reporting & Log all actions in a tamper-proof system; provide rectification reports for audits. \\
\hline
RQ8 & User Notification & Notify data subject when changes are processed, including details of corrections. \\
\hline
RQ9 & Third-Party Rectification Data Sharing & Notify third parties about corrections if inaccurate data was involved in their operations. \\
\hline
\end{tabularx}
\end{table*}

This article will be directly relevant to the BraneHub portal, if it is deployed for public use. A data provider should be able to rectify the data submitted via the questionnaire in Table~\ref{tab:participant-interface}. If the participant updates the location or region, this update can have an implication for the workflow that the researcher must address, e.g., by excluding the participant from the study or, possibly, by adapting the policies to process the data in the updated region. 

\begin{figure*}
    \centering
    \includegraphics[width=1.0\textwidth]{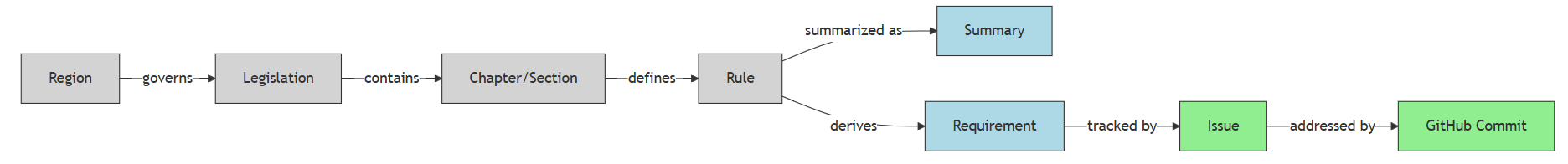}
    \caption{Compliance rule management via GutHub issues}
    \label{fig:compliance-github}
\end{figure*}

Requirements RQ1-RQ9 cannot be addressed exclusively at the application level. Hence, we propose to link the selected rules originating from legislation to the relevant code snippets implementing the operational support for such requirements via the GitHub issues as shown in Figure~\ref{fig:compliance-github}. Legal requirements relevant for the FDP participant region are split into chunks (chapters, sections, and summaries to capture the overall requirements in a concise form), vectorized, and interpreted as rules and system requirements. Developers can refer to rules in code comments or associate code commits with GitHub issues that connect system requirements with regulatory document articles.          

\subsection{Runtime Access Control Policies}

The collaborative research on sensitive data strongly depends on many factors, including \emph{purpose}. Access policies are critical to control who can access data for processing, under what conditions, and for how long. The theory of access control modeling is well established and offers off-the-shelf principles and guidelines currently used in nearly every software system:

\begin{itemize}
    \item \emph{Role-Based Access Control (RBAC)}~\cite{sandhu1996rbac}: Access is granted based on the individual’s role (e.g., doctor, compliance officer, principle investigator). 
    \item \emph{Attribute-Based Access Control (ABAC)}~\cite{hu2015abac}: Access depends on data attributes (e.g., patients' age, treatment time, location). 
    \item \emph{Consent-Based Access}: Data use must align with person's consent, study-specific permissions, data use agreements, or IRB clauses.
    \item \emph{Purpose Limitation}: Data may be accessed only for explicitly approved purposes (e.g., IRB-approved questions only).
    \item \emph{Time-Limited Access}~\cite{Bertino2000TemporalAB}: Access is granted for a defined period (e.g., duration of a trial phase or a specific study window). This implies synchronization with organizational structures to ensure that consultants, temporary staff, and short-term collaborators do not retain access beyond their time in an assigned role.
    \item \emph{Data Minimization and Least Privilege}: Users should only see the minimum data necessary for their task. FDP data providers should  grant access to already de-identified, pseudonymized, or aggregated data to reduce risk. However, FDP participants, primarily, project owner, should further safeguard the data by removing or masking  identifiable elements.
    \item \emph{Audit and Logging}: All data access events must be logged to meet regulatory and ethical standards~\cite{varkey2021clinical_ethics}.
\end{itemize}

RBAC is simple to implement but less flexible; ABAC allows dynamic, context-aware access, including consent status, data sensitivity, and temporal or jurisdictional constraints, but puts higher technical demands on workflow designer. Table~\ref{tab:ABAC} and Table~\ref{tab:RBAC} provide role and policy specifications for our running scenario, which translate to the Rego/OPA policy shown in Listing~\ref{lst:rbac}. While this is not a trivial setup for a domain scientist not experienced with access policy definition guidelines and technical solutions, it is much easier with the help of the AI assistant exposed to the project context via the structured information collected in BraneHub. 

\begin{table*}[ht]
\centering
\scriptsize
\caption{Example RBAC roles and permissions for pharma--hospital data collaboration}
\label{tab:RBAC}
\setlength{\tabcolsep}{4pt}
\renewcommand{\arraystretch}{1.3}
\begin{tabular}{|>{\raggedright\arraybackslash}p{3.0cm}
                |>{\raggedright\arraybackslash}p{3.8cm}
                |>{\raggedright\arraybackslash}p{2.4cm}
                |>{\raggedright\arraybackslash}p{3.4cm}|}
\hline
\textbf{Role} & \textbf{Accessible Data} & \textbf{Permitted Actions} & \textbf{Notes} \\
\hline
\texttt{EpidemiologyAnalyst}
  & De-identified cohort statistics,\newline prevalence and incidence records
  & Read, Analyze, Export
  & Early-stage demand assessment; no access to individual records \\
\hline
\texttt{ClinicalResearcher}
  & Pseudonymized EHR records,\newline lab results, imaging data
  & Read, Analyze,\newline Annotate
  & Supports PK/PD modelling, AI methods, and digital twin evaluation \\
\hline
\texttt{DataManager}
  & Full pseudonymized EHR records
  & Read, Transform,\newline Anonymize
  & Prepares and curates datasets for R\&D pipelines \\
\hline
\texttt{TrialCoordinator}
  & Eligibility-filtered patient identifiers
  & Read, Contact patient
  & Restricted to recruitment window; requires ethics approval \\
\hline
\texttt{EthicsOfficer}
  & Access logs, consent records,\newline audit metadata
  & Review, Audit
  & Monitors regulatory compliance and consent adherence \\
\hline
\end{tabular}
\end{table*}

\begin{table*}[ht]
\centering
\scriptsize
\caption{Example ABAC policies for pharma--hospital data collaboration}
\label{tab:ABAC}
\setlength{\tabcolsep}{4pt}
\renewcommand{\arraystretch}{1.3}
\begin{tabular}{|>{\raggedright\arraybackslash}p{1.8cm}
                |>{\raggedright\arraybackslash}p{2.8cm}
                |>{\raggedright\arraybackslash}p{3.4cm}
                |>{\raggedright\arraybackslash}p{2.8cm}
                |>{\centering\arraybackslash}p{1.4cm}|}
\hline
\textbf{Policy} & \textbf{User Attributes} & \textbf{Resource Attributes} & \textbf{Context Attributes} & \textbf{Action} \\
\hline
Cohort prevalence access
  & \texttt{role} = EpidemiologyAnalyst
  & \texttt{data\_type} = aggregate\_stats,\newline \texttt{condition} = target\_condition
  & \texttt{location} = EU,\newline \texttt{consent} = true
  & Read,\newline Export \\
\hline
PK/PD model data access
  & \texttt{role} = ClinicalResearcher
  & \texttt{data\_type} = EHR\_pseudonymized,\newline \texttt{imaging} = true
  & \texttt{location} = EU,\newline \texttt{data\_anonymized} = true
  & Read,\newline Analyze \\
\hline
Dataset preparation
  & \texttt{role} = DataManager
  & \texttt{data\_type} = EHR\_full,\newline \texttt{pseudonymized} = true
  & \texttt{data\_anonymized} = true
  & Read,\newline Transform \\
\hline
Trial recruitment
  & \texttt{role} = TrialCoordinator,\newline \texttt{ethics\_approved} = true
  & \texttt{eligibility} = true,\newline \texttt{data\_type} = identifiable
  & \texttt{current\_date} $\in$ \texttt{trial\_window}
  & Contact\newline patient \\
\hline
Compliance audit
  & \texttt{role} = EthicsOfficer
  & \texttt{data\_type} = access\_logs,\newline \texttt{consent} = true
  & \texttt{audit\_period} = active
  & Review,\newline Audit \\
\hline
\end{tabular}
\end{table*}

\begin{lstlisting}[language=Rego, caption={RBAC policy in Rego/OPA},
label={lst:rbac}]
package pharma.rbac

# Role permissions mapping
role_permissions := {
    "EpidemiologyAnalyst": {
        "resources": {"aggregate_stats"}, "actions":   {"read", "analyze", "export"}},
    "ClinicalResearcher": {
        "resources": {"ehr_pseudonymized"}, "actions":   {"read", "analyze", "annotate"}},
    "DataManager": {
        "resources": {"ehr_full"}, "actions":   {"read", "transform", "anonymize"}},
    "TrialCoordinator": {
        "resources": {"identifiable_records"}, "actions":   {"read", "contact_patient"}},
    "EthicsOfficer": {
        "resources": {"access_logs", "consent_records"}, "actions":   {"review", "audit"}}
}

# Input: { "role": "ClinicalResearcher", "action": "read", "resource": "ehr_pseudonymized" }
default allow := false

allow if {
    perms := role_permissions[input.role]
    input.resource in perms.resources
    input.action in perms.actions
}
\end{lstlisting}

Support for the development of Rego policies within BraneHub can be introduced incrementally. Currently, the portal provides a viewer with a server-side syntax checking using the OPA CLI (opa check), which validates policy correctness without editor integration. The OPA ecosystem provides an online playground to edit and test policies~\cite{opa_playground}, the tool is embedded to the BraneHub via the iFrame. A slightly richer approach would add server-side linting with the Regal CLI~\cite{openpolicyagent_regal}, enabling detection of best-practice violations and integration with the AI assistant to validate auto-generated policies. More advanced integration can use a browser-based code editor with custom Rego language configuration for highlighting and error markers. The most comprehensive solution could combine an editor with the Regal language server via the language server protocol (LSP)~\cite{regal-language-server}, offering IDE-level features such as real-time diagnostics, semantic validation, code navigation, and auto-completion directly in the browser. 

Policy management support can further be strengthen by the extension of BraneHub with formal verification~\cite{temporal_rbac_alloy_2014}, model checking~\cite{model_checking_access_control_2023}, automated simulation, and graphical visualization~\cite{graph_framework_access_control_2021} of collective FDP network data access specifications. Such options would help at the design stage to make sure data privacy is not compromised by, e.g., unauthorized access after project end or by an individual with insufficient permissions. In the next section, we outline the formal ground for formal access rule validation in FDP networks.     

\section{Spatio-Temporal Purpose-Aware RBAC Graph}

We define a \emph{Spatio-Temporal Purpose-Aware RBAC Graph (STP-RBACG)}
as a formal model for enforcing temporal, regional, and privacy-aware
constraints over distributed data access in the FDP network. Let 
\begin{itemize}
  \item $R = \{R_M, R_0, \dots, R_k\}$ be a finite set of participant roles,
  \item $O = \{d_0, \dots, d_n\}$ be a finite set of data objects,
  \item $\mathcal{L}$ be a finite set of geographic or jurisdictional labels,
  \item $\mathbb{T}$ be a totally ordered time domain,
  \item $\mathcal{P}$ be a finite set of processing purposes.
\end{itemize}

An \emph{STP-RBACG instance} is the tuple:
\[
  \mathcal{T}_{\mathrm{STP}} =
  (R,\, O,\, \mathcal{L},\, \mathcal{P},\,
   PA,\, \tau,\, \rho_R,\, \rho_O,\, \Gamma,\, \Pi,\, \Delta)
\]
where:
\begin{itemize}
  \item $PA \subseteq R \times O$ is the role-to-object permission assignment,
  \item $\tau : PA \to \mathbb{T} \times \mathbb{T}$ assigns each assigned
        pair $(r,o)$ a validity interval
        $\tau(r,o) = (t^s_{(r,o)},\, t^e_{(r,o)})$
        with $t^s_{(r,o)} \leq t^e_{(r,o)}$,
  \item $\rho_R : R \to 2^{\mathcal{L}}$ maps each role to the set of
        jurisdictional labels of the participant holding that role,
  \item $\rho_O : O \to 2^{\mathcal{L}}$ maps each data object to the set
        of jurisdictional labels of its storage location(s),
  \item $\Gamma \subseteq \mathcal{L} \times \mathcal{L}$ is the
        \emph{cross-jurisdiction access relation}: $(l_r, l_o) \in \Gamma$
        means that a participant in jurisdiction $l_r$ is permitted to
        access data governed by jurisdiction $l_o$.
        $\Gamma$ is not assumed to be symmetric, reflecting asymmetries
        in adequacy decisions and bilateral data-sharing agreements,
  \item $\Pi : PA \to 2^{\mathcal{P}}$ maps each permission pair $(r,o)$
        to the set of purposes for which role $r$ may process object $o$.
        Defining $\Pi$ on $PA$ rather than on $R$ alone captures
        object-specific purpose restrictions (e.g.\ an imaging dataset
        may be accessible to a data manager for anonymisation but not
        for export),
  \item $\Delta : \mathcal{P} \to 2^{O}$ maps each purpose to its minimal
        required set of data objects.
\end{itemize}

The \emph{temporal authorization graph} is the directed bipartite graph:
\[
  G = (V,\, E,\, \tau), \qquad
  V = R \cup O, \qquad E = PA,
\]
with each edge $e = (r,o) \in E$ carrying the activation interval
$\tau(e) = (t^s_e,\, t^e_e)$.
The \emph{edge activation function} is:
\[
  \chi_e(t) =
  \begin{cases}
    1 & t^s_e \leq t \leq t^e_e, \\
    0 & \text{otherwise.}
  \end{cases}
\]
The \emph{time-indexed graph} at time $t$ retains only currently active edges:
\[
  G(t) = (V,\, E_t), \qquad
  E_t = \bigl\{ e \in E \mid \chi_e(t) = 1 \bigr\}.
\]
$G(t)$ governs which role--object pairs carry a live temporal permission at $t$,
but temporal activation alone is not sufficient for access authorization:
the spatial, purpose, and minimization constraints below must also be satisfied.

An \emph{access request} is a tuple $(r, o, p, t) \in R \times O \times \mathcal{P} \times \mathbb{T}$.
The following constraints must hold system-wide in any FDP network instance.

\begin{definition}[Region Compatibility]
Role $r$ is \emph{region-compatible} with object $o$ iff the
jurisdiction of the participant and the jurisdiction governing the
data are connected by an admissible cross-jurisdiction pair:
\[
  \mathit{RegionAllowed}(r, o)
  \;\iff\;
  \exists\, l_r \in \rho_R(r),\;
  \exists\, l_o \in \rho_O(o) :
  (l_r, l_o) \in \Gamma.
\]
Because $\Gamma$ is not symmetric, $\mathit{RegionAllowed}(r,o)$ does not
imply $\mathit{RegionAllowed}(r',o)$ for a role $r'$ in the converse
jurisdiction, nor does it imply $\mathit{RegionAllowed}(r,o')$ for an
object $o'$ governed by a different jurisdiction.
\end{definition}

\begin{definition}[Purpose Limitation]
An access request $(r,o,p,t)$ satisfies \emph{purpose limitation} iff
the stated purpose is within the set of purposes permitted for the
specific role--object pair:
\[
  \mathit{PurposeLimited}(r, o, p)
  \;\iff\;
  p \in \Pi(r, o).
\]
\end{definition}

\begin{definition}[Data Minimisation]
A requested object set $O' \subseteq O$ satisfies \emph{data minimisation}
for purpose $p$ iff every requested object belongs to the minimal
required set for that purpose:
\[
  \mathit{MinData}(O', p)
  \;\iff\;
  O' \subseteq \Delta(p).
\]
For a single object $o$, this reduces to $o \in \Delta(p)$.
A request for any object outside $\Delta(p)$ constitutes a minimisation
violation regardless of the status of the remaining requested objects.
\end{definition}

The \emph{active spatio-temporal edge set} at time $t$ encodes the
temporal and spatial constraints statically:
\[
  E_t^{\,\mathrm{ST}} =
  \bigl\{
    (r,o) \in PA
    \;\big|\;
    \chi_{(r,o)}(t) = 1
    \;\land\;
    \mathit{RegionAllowed}(r,o)
  \bigr\},
\]
yielding the \emph{active spatio-temporal graph}:
\[
  G^{\mathrm{ST}}(t) = (R \cup O,\; E_t^{\,\mathrm{ST}}).
\]
Purpose limitation and data minimization are \emph{not} encoded in the
graph structure: they depend on the purpose $p$ supplied at query time,
which varies per request and cannot be precomputed into a static edge
set without creating a separate graph instance for every purpose.
Therefore, these two constraints are evaluated per-request at decision time, against $G^{\mathrm{ST}}(t)$ as the prefiltered candidate set.

\begin{definition}[Spatio-Temporal Purpose-Aware Authorization]
An access request $(r, o, p, t)$ is \emph{authorized} iff all five
conditions hold jointly:
\[
  \begin{aligned}
  \mathit{Auth}(r,o,p,t)
  \;\iff\;
  \underbrace{(r,o) \in PA}_{\text{assignment}}
  \;\land\;
  \underbrace{\chi_{(r,o)}(t) = 1}_{\text{temporal}}
  \;\land\;
  \underbrace{\mathit{RegionAllowed}(r,o)}_{\text{spatial}} \\
  \;\land\;
  \underbrace{\mathit{PurposeLimited}(r,o,p)}_{\text{purpose}}
  \;\land\;
  \underbrace{\mathit{MinData}(\{o\},p)}_{\text{minimisation}}.
  \end{aligned}
\]

The condition $(r,o) \in PA$ is stated explicitly: $\chi_{(r,o)}$ is
defined on $PA$ but an authorization predicate must reject requests
for pairs not in the assignment, not merely treat them as inactive.
\end{definition}

\begin{definition}[Spatio-Temporal Privacy Compliance]
A system satisfies \emph{spatio-temporal privacy compliance} iff every
authorized request corresponds to legally valid processing:
\[
  \forall\, r \in R,\;
  o \in O,\;
  p \in \mathcal{P},\;
  t \in \mathbb{T} :
  \quad
  \mathit{Auth}(r, o, p, t)
  \;\Rightarrow\;
  \mathit{LegallyValid}(r, o, p, t).
\]
Equivalently, by contrapositive, any access that is not legally valid
must not be authorized. A \emph{compliance violation} occurs when
$(r,o) \in PA$, $\chi_{(r,o)}(t) = 1$ (the temporal gate passes), but
at least one of $\mathit{RegionAllowed}$, $\mathit{PurposeLimited}$, or
$\mathit{MinData}$ fails — the system grants a time-valid edge without
satisfying the substantive constraints.
\end{definition}

This formalization captures network-level contextual constraints spanning time and jurisdictions and enables distributed enforcement across the FDP network. Policy correctness emerges from a coordinated system-wide specification: neither data provider nor consumer alone can guarantee both protection against unauthorized access and support for legitimate workflows. Temporal and spatial constraints are precomputed in $G^{\mathrm{ST}}(t)$ and enforced at the coordinator layer, filtering requests before they reach data providers. Purpose limitation and data minimisation, which require the full request context, are evaluated per-request at the worker nodes in the data provider domain, ensuring that each role-object access satisfies the purpose-specific minimal data set. Researcher nodes issue requests specifying the intended purpose, which are then authorized only if all constraints pass, balancing efficiency, flexibility, and compliance.

\section{STP-RBACG Instance for the Running Scenario}

We illustrate STP-RBACG with a concrete scenario and then demonstrate
how access decisions are made. We assume that temporal intervals, jurisdictions, purposes, and minimal data sets are collected via the BraneHub and defined for each role-object pair.

\paragraph{Roles $\mathcal{R}$:} 
$r_1$ — EpidemiologyAnalyst, $r_2$ — ClinicalResearcher, $r_3$ — DataManager, 
$r_4$ — TrialCoordinator, $r_5$ — EthicsOfficer.

\paragraph{Data Objects $\mathcal{O}$:} 
$d_1$ — Cohort stats, $d_2$ — Pseudonymized EHR, $d_3$ — Pediatric imaging, 
$d_4$ — Lab results, $d_5$ — Identifiable patient records, $d_6$ — Consent/audit logs.

\paragraph{Purposes $\mathcal{P}$:} 
$p_1$ — Epidemiological feasibility, $p_2$ — PK/PD modeling \& digital twin, 
$p_3$ — Dataset preparation \& anonymization, $p_4$ — Clinical recruitment, $p_5$ — Regulatory audit.

\paragraph{Time Domain $\mathbb{T}$ (months):} 
Feasibility $[0,12]$, Development $[12,48]$, Recruitment $[36,54]$, Audit $[0,60]$.

\paragraph{Jurisdictions $\mathcal{L}$:} $\{\text{NL},\text{DE},\text{UK},\text{EU},\text{US}\}$.

\paragraph{Permission Assignment $PA$ and Intervals $\tau$}

\begin{center}
\begin{tabular}{cllcccl}
\toprule
Edge & Role & Object & Interval & Role Juris. & Object Juris. & Purpose / $\Delta$ \\
\midrule
$e_1$ & $r_1$ & $d_1$ & $[0,12]$ & NL & NL,DE & $p_1$ / $\{d_1\}$ \\
$e_2$ & $r_2$ & $d_2$ & $[12,48]$ & NL,DE & NL,DE & $p_2$ / $\{d_2,d_3,d_4\}$ \\
$e_3$ & $r_2$ & $d_3$ & $[12,48]$ & NL,DE & NL,DE & $p_2$ / $\{d_2,d_3,d_4\}$ \\
$e_4$ & $r_2$ & $d_4$ & $[12,48]$ & NL,DE & NL,DE & $p_2$ / $\{d_2,d_3,d_4\}$ \\
$e_5$ & $r_3$ & $d_2$ & $[0,48]$ & NL & NL,DE & $p_3$ / $\{d_2,d_3\}$ \\
$e_6$ & $r_3$ & $d_3$ & $[0,48]$ & NL & NL,DE & $p_3$ / $\{d_2,d_3\}$ \\
$e_7$ & $r_4$ & $d_5$ & $[36,54]$ & NL,DE,UK & NL,DE,UK & $p_4$ / $\{d_5\}$ \\
$e_8$ & $r_5$ & $d_6$ & $[0,60]$ & NL & NL,DE & $p_5$ / $\{d_6\}$ \\
\bottomrule
\end{tabular}
\end{center}

\paragraph{Cross-Jurisdiction Relation $\Gamma$:} 
$\{(\text{NL},\text{NL}),(\text{NL},\text{DE}),(\text{DE},\text{NL}),(\text{DE},\text{DE}),(\text{NL},\text{UK}),(\text{UK},\text{NL})\}$.

\paragraph{Example Authorizations:}
\begin{itemize}[nosep]
\item \textbf{Authorized:} $\mathit{Auth}(r_2,d_3,p_2,20)$ — all checks pass (temporal, jurisdiction, purpose, minimal data).
\item \textbf{Denied (temporal):} $\mathit{Auth}(r_4,d_5,p_4,30)$ — recruitment not open ($t\notin[36,54]$).
\item \textbf{Denied (assignment):} $\mathit{Auth}(r_2,d_5,p_2,20)$ — $(r_2,d_5)\notin PA$, structural check fails.
\item \textbf{Denied (purpose):} $\mathit{Auth}(r_3,d_3,p_2,20)$ — $(r_3,d_3)\in PA$ but $p_2\notin \Pi(r_3,d_3)$; only $p_3$ allowed.
\end{itemize}

\section{Discussion}
\label{sect:discussion}
In this section, we discuss the readiness of the proposed framework and outline the main challenges, including evaluation strategies and production-level deployment.

The overarching goal of the framework is to support the setup, operation, and long-term management of FDP networks in real-world settings. To demonstrate fitness for purpose, the framework must enable domain scientists to execute the full project lifecycle: defining workflows, specifying participant-specific and context-dependent access policies, deploying these workflows across federated nodes, and securely consuming results derived from protected data. A key requirement is that this process remains feasible without deep expertise in distributed systems, security engineering, or formal policy languages. 

At this stage, the system has been validated at the level of individual components through unit testing. Specifically, we evaluated document ingestion and retrieval within the RAG module, the translation of semi-structured participant and dataset descriptions into OPA/Rego policies, and the decision-making capabilities of both LLM-based and OPA/Rego-based onboarding assistants. The latter were tested for their ability to accept or reject onboarding requests and to generate clear, context-aware explanations supporting their decisions.

The AI assistant plays a central role in lowering the  barrier for FDP network deployment by providing the following capabilities:
\begin{enumerate}
\item discovery of relevant legal and regulatory requirements given a project context,
\item translation of requirements into machine-actionable artifacts (e.g., OPA/Rego or eFLINT policies, system requirements, and code-linked documentation such as comments or GitHub issues),
\item explanation of formal policies in domain-specific, user-friendly language,
\item advisory support for compliance-related questions concerning data and workflows,
\item assistance in identifying and resolving mismatches in data formats and semantics,
\item assisting with scripts to package and distribute policies across federated nodes.
\end{enumerate}
Consequently, relevant research questions include: (i) how effectively LLMs identify compliance requirements for a given FDP context; (ii) how accurately legal and organizational requirements are translated into formal rules and back; and (iii) how LLM-based compliance decisions compare to expert human judgments in terms of false positive and false negative rates. These decisions may be produced directly by LLMs or mediated through policy engines such as OPA/Rego, with or without additional contextual grounding via RAG.
While operational guidelines exist on how to extract access rights and obligations directly from regulation texts~\cite{regRules}, a domain scientist cannot be expected to perform such a transformation from scratch; the FDP setup platform should guide the user. 

To assess rule generation and translation quality, complementary metrics exist, including: (i) \emph{semantic fidelity}, measuring alignment with the source normative text; (ii) \emph{legal correctness}, ensuring preservation of obligations, prohibitions, permissions, and exceptions; (iii) \emph{grounding and factuality}, verifying support by retrieved evidence; and (iv) \emph{retrieval quality}, evaluated using precision, recall, mean reciprocal rank (MRR), and grounding rate. Some existing  publications provide insights on the effectiveness of LLM-based translation of natural language reqirements to formal  specifications~\cite{ma2025NL2LTL,JudgeBert}. Our framework does not offer innovative solutions on this aspect, we merely rely on the integration of general purpose (or fine-tuned for legal tasks~\cite{Bhatkar2024}) LLMs to accelerate translation of regulatory constraints to enforceable policies within FDP networks. The comparative evaluation of LLM- and OPA/Rego-based onboarding advice against expert decisions (for patient cohort selection and clinical trial recruitment) is ongoing and will be reported in future work).

Beyond correctness, usability and maintainability are critical evaluation dimensions. We should assess whether domain scientists find the framework intuitive and supportive, whether policy provenance and decision chains remain traceable from legal sources to concrete data access events, and whether the introduction of autonomous agents (e.g., for schema conversion or record mapping)~\cite{neubauer2025ai} introduces new security or compliance risks. Additional evaluation criteria include the expressiveness of the policy language from the user perspective and the ease of integrating evolving policies into deployed workflows.

\section{Related Work}
\label{sect:related-work}

To address the complexity of global data protection regulations, several comprehensive resources have emerged. DataGuidance~\cite{dataguidance2025} provides regulatory intelligence across 300+ jurisdictions, offering extensive legal analysis, comparative tools, and AI-supported research integrated into OneTrust’s Privacy Automation platform. DLA Piper’s Data Protection Laws of the World~\cite{dlapiper2025dataprotection} delivers an interactive comparative overview of privacy regimes in over 160 jurisdictions, while the International Association of Privacy Professionals (IAPP)~\cite{iapp2025resources} supports a global community of 80,000+ members with research, legislative tracking, enforcement databases, and certification programs. Replicating such resources in our academic effort is neither feasible nor necessary. We integrate an LLM/RAG AI-assistant  to illustrate its integration with the rest of the architectural solution. The vector space is extended dynamically based on a new project requirements or triggered by onboarding requests from a certain jurisdiction.   

A key barrier to real-world deployment of federated infrastructures is their inability to operationalize legal requirements as enforceable system behavior. While FL frameworks increasingly support privacy-enhancing techniques such as differential privacy and secure aggregation \cite{geyerdp,shokri2015privacypreserving,truex2019hybrid}, compliance with the GDPR and related legal regimes is typically treated as an \emph{external} organizational duty rather than a first-class computational concern. Existing platforms largely protect data confidentiality and integrity, yet provide no explicit guarantees regarding legal rights such as rectification, erasure, data minimisation, purpose limitation, or processing traceability. These duties must ultimately be realized at the system and workflow layers.

\subsection{Compliance Challenges in Federated Platforms}
Several recent FL deployments in healthcare
\cite{brisimi2018federated,rieke2020future,kaissis2020secure,kurtz2021federated}
demonstrate clinical feasibility but delegate compliance reasoning to institutional lawyers or local policy documents rather than embedding it into the orchestration layer. This results in fragmented and manual governance processes \cite{quinn2022federated}. Even advanced research platforms such as Fed-BioMed \cite{Cremonesi2025FedBioMed} treat legal compliance as an adjunct responsibility: the platform enables secure computation but does not provide a computable representation of legal obligations nor a traceable link between legislation and configuration-level enforcement. 

The fragility of this assumption is illustrated by Matte et al.~\cite{matte2020cookie} who show that public-sector digital services maintained by experienced software teams frequently fail to comply with their own stated privacy guarantees. This highlights that \emph{documentation-level} compliance is insufficient; verifiable, system-level enforcement is required. More broadly, the variability and frequent revision of legal obligations across jurisdictions force organisations to maintain multiple software variants, increasing operational burden and discouraging federated collaboration.

\subsection{Formal Specification Approaches}
A long-standing line of work explores the use of formal policy languages, normative DSLs, and process modeling to encode compliance rules as machine-interpretable artifacts. Early works by Schumm et~al.~\cite{schumm2010} introduced reusable \emph{process fragments} as declarative compliance controls integrated into business process models. Zasada et~al.~\cite{zasada23} provided a comparative evaluation of workflow-compliance languages, identifying a tension between expressiveness and usability. More recent frameworks such as eFLINT \cite{VANBINSBERGEN2022140} model legal and organizational norms as executable specifications from which regulatory services can be generated. While such DSL-based approaches offer fine-grained control and principled separation of concerns, they are labor-intensive, demand specialized training, and remain weakly integrated with continuous development tooling (e.g., no native coupling between legal artifacts and the version-controlled implementation of workflows).

JustAct~\cite{EsterhuyseMVB24} and JustAct+\cite{JustActPlus2025} present decentralized frameworks for multi‑agent systems that enable agents to justify and audit their actions against inter‑organizational policies. The approach leverages logic‑based policy fragments that agents autonomously share and assemble, supporting accountability and verifiability of data processing actions in privacy‑sensitive domains such as healthcare. The framework is demonstrated through a case study reproducing the Brane medical data processing system.

\subsection{LLM-based Compliance Automation}
LLMs have recently been explored as a bridge between natural-language regulation and automated technical governance. Hassani et~al.\ \cite{hassani2024enhancing,hassani2024rethinkinglegalcomplianceautomation} show that LLMs improve contextual interpretation of GDPR provisions and produce more faithful alignments between regulatory clauses and obligations than rule-based extraction. Kim and Min \cite{kim2024ragqaragintegratinggenerative} extend this by applying RAG to pharmaceutical regulation, reducing hallucination by grounding responses in policy documents. Garza et~al.\ \cite{garza2024privcomp} combine LLM reasoning with knowledge-graph semantics, while Li and Maiti \cite{li2025applying} propose continuous monitoring pipelines for industrial compliance. Further research explores multi-agent LLM simulations of regulatory negotiation \cite{han2024regulatormanufactureraiagentsmodeling} and real-time dynamic policy adaptation \cite{sobkowski2024dawn}. However, these approaches operate primarily in an \emph{advisory} capacity: they extract, explain, or recommend obligations but do not integrate them into the technical lifecycle of workflow deployment or provenance tracking.

\subsection{Integrating Compliance with Federated Workflows}
Our work builds on the aforementioned developments by integrating compliance interpretation with workflow execution. Legal documents are embedded into a vector store to enable retrieval-augmented translation of regulatory text into system-level requirements that are contextualized by the structure of the federated collaboration. These requirements are decomposed into actionable development artifacts (e.g., GitHub issues) and linked to code fragments through commit metadata, establishing a verifiable traceability chain from \emph{legislation $\rightarrow$ derived requirement $\rightarrow$ issue $\rightarrow$ code $\rightarrow$ deployed workflow}. In contrast to prior LLM-based solutions, which remain detached from runtime orchestration, this approach enables \emph{lifecycle-aware} compliance: as new participants or jurisdictions join, newly applicable obligations are surfaced and corresponding implementation changes can be invoked.

Operationally, workflows are executed as VNFs deployed within Brane containers, which allows computation to move to the data rather than relocating data to a central orchestrator. This architectural property supports data sovereignty and jurisdictional localization.

\section{Conclusions and Future Work}
\label{sect:conclusions}

This paper presented a framework for compliance-aware FDP designed to lower the barrier to deploying federated analytics in regulated domains. The framework addresses a gap that existing FDP platforms leave largely unresolved: the systematic translation of legal and organizational requirements into machine-actionable, runtime-enforceable artifacts, without requiring domain scientists to engage directly with formal policy languages or distributed systems engineering.

The core contributions are threefold. First, we introduced BraneHub, a coordination platform that supports the full pre-integration lifecycle — project registration, participant onboarding, structured compliance questionnaires, and AI-assisted policy authoring — and integrates with the Brane execution framework for runtime access control. Second, we demonstrated that LLM- and RAG-assisted components can support policy discovery, requirement-to-policy translation, onboarding decision support, and compliance explanation, complementing formal OPA/Rego evaluation where structured rule matching alone is limiting. Third, we proposed a Spatio-Temporal Purpose-Aware RBAC Graph (STP-RBACG) as a formal foundation for reasoning about temporal, jurisdictional, and purpose-based access constraints in FDP networks.

Together, these contributions operationalize compliance-by-design across the project lifecycle, from legislation to derived requirements, policy specification, workflow deployment, and traceability via linked development artifacts. The prototype implementation validates individual components; full end-to-end evaluation across realistic multi-institutional deployments, quantitative assessment of LLM-based compliance decisions against expert judgements, and strengthened tooling for long-term monitoring and audit remain priorities for future work.

\bibliographystyle{spbasic}
\bibliography{main}

\end{document}